\documentclass[a4paper,11pt]{article}
\pdfoutput=1 % if your are submitting a pdflatex (i.e. if you have
\usepackage{jinstpub} % for details on the use of the package, please
\graphicspath{{images/}}

\title{\boldmath From 3D to 5D tracking: SMX ASIC-based Double-Sided Micro-Strip detectors for comprehensive space, time, and energy measurements}

\author[a,1]{M. Teklishyn,\note{Corresponding author.}}
\author[a]{A. Rodr\'iguez Rodr\'iguez,}
\author[c]{K. Agarwal,}
\author[a,b]{M. Bajdel,}
\author[b,a]{L.\,M. Collazo S\'anchez,}
\author[a]{U. Frankenfeld,}
\author[a]{J.\,M. Heuser,}
\author[a]{J. Lehnert,}
\author[a,c]{S. Mehta,}
\author[a,b]{D. Rodr\'iguez Garc\'es,}
\author[a,b]{D.\,A. Ram\'irez Zald\'ivar,}
\author[a]{C.\,J. Schmidt,}
\author[a,c]{H.\,R. Schmidt,}
\author[a,b]{A. Toia}

\affiliation[a]{GSI Helmholtzzentrum für Schwerionenforschung GmbH, Darmstadt, Germany}
\affiliation[b]{Goethe-Universität Frankfurt, Germany}
\affiliation[c]{Eberhard Karls Universität T\"{u}bingen, Germany}

\emailAdd{m.teklishyn@gsi.de}

\abstract{We present the recent development of a lightweight detector capable of accurate spatial, timing, and amplitude resolution of charged particles. The technology is based on double-sided double-metal p+\,--\,n\,--\,n+ micro-strip silicon sensors, ultra-light long aluminum-polyimide micro-cables for the analogue signal transfer, and a custom-developed SMX read-out ASIC capable of measurement of the time ($\Delta t \lesssim 5 \,\mathrm{ns}$) and amplitude. Dense detector integration enables a material budget $>0.3\,\% X_0$. A sophisticated powering and grounding scheme keeps the noise under control.

In addition to its primary application in Silicon Tracking System of the future CBM experiment in Darmstadt, our detector will be utilized in other research applications.}


\proceeding{Topical Workshop on Electronics for Particle Physics\\
  2 – 6 October 2023 \\
  Geremeas, Sardinia, Italy}

\notoc

\begin{document}
\maketitle
\flushbottom

\section{Introduction: Silicon Tracking System of the CBM experiment}
\label{sec:intro}

% \subsection{Silicon Tracking System of the CBM experiment}

The Silicon Tracking System (STS) of the Compressed Baryonic Mater (CBM) experiment is the core application and main motivation for the detector technology described in this paper.

CBM is a fixed-target heavy-ion experiment of the future Facility for Antiproton and Ion Research (FAIR) complex in Darmstadt, Germany; it is dedicated to study the strongly interacting matter under extreme conditions \cite{Agarwal2023}.

The CBM detector is a single-arm forward spectrometer capable of collecting data in a free-streaming mode to process the unprecedented beam-target interaction rates of up to $10\,\mathrm{MHz}$. The detector will utilise the heavy-ion and proton beams from the SIS-100 accelerator at energies of up to $11\,A\mathrm{GeV}$ and $29\,\mathrm{GeV}$ \cite{Durante2019}.

STS is the core tracking detector of CBM. Its 8 tracking stations (876 detector modules, total silicon area about $4\,\mathrm{m^2}$) will be placed $30-100\,\mathrm{cm}$ downstream of the target in the aperture of the $1\,\mathrm{T\!\cdot\!m}$ superconductive dipole.  The primary goals of the STS are tracking of charged particles ($\lesssim700$ tracks in the Au+Au central collision), momentum determination with $\delta p/p \lesssim 1.5\%$ and secondary vertex reconstruction. For these tasks STS requires a position resolution better than  $30\,\mathrm{\mu m}$ in the bending plane, a good time resolution (in the order of $5 - 10\,\mathrm{ns}$) and a material budget within $0.3\%-1.4\%X_0$ per tracking station \cite{Lymanets2019, Lavrik2019, Schmidt2019, Rodriguez2019}.

\section{Micro-strip Silicon module: the fundamental functioning block of STS}

A key component of the STS detector module is the custom STS-MUCH-XYTER (SMX) Application-Specific Integrated Circuit (ASIC) developed for the front-end electronics of the STS and Muon Chamber (MUCH) detectors of CBM. It features 128 channels each capable of simultaneous measurement of the signal amplitude and time with 5-bit ADC (dynamic range $<15\,\mathrm{fC}$ for the STS and $<100\,\mathrm{fC}$ for MUCH modes) and 14-bit TDC ($\Delta T_\text{LSB}=3.125\,\mathrm{ns}$) in a free-streaming mode \cite{Kasinski2018}.

\begin{figure}[htbp]
\centering
\includegraphics[width=.50\textwidth]{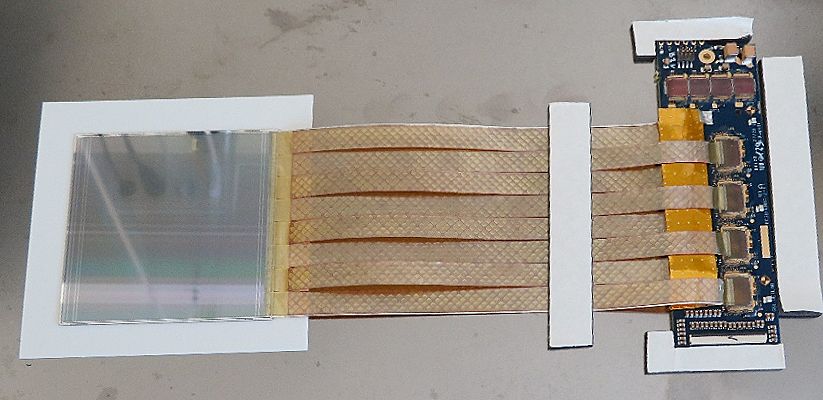}
\caption{\label{fig:photo_module}
Picture of the partially assembled E16-104 module with $62\times62\,\mathrm{mm^2}$ sensor and $121\,\mathrm{mm}$ micro-cable. Micro-cable shielding, aluminium cooling fin and return path circuit wire are missing.}
\end{figure}

An STS detector module consists of a $320\,\mathrm{\mu m}$ thick, $62\,\mathrm{mm}$ wide p+\,--\,n\,--\,n+ double-sided double-metal (DSDM) silicon microstrip sensor of four versions: $22\,\mathrm{mm}$, $42\,\mathrm{mm}$, $62\,\mathrm{mm}$ and $124\,\mathrm{mm}$ long. The sensors are manufactured by Hamamatsu Photonics \cite{Hamamatsu} using high-resistance silicon wafers: typical dark current is $I_d<40\,\mathrm{nA/cm^2}$ at $20^\circ\mathrm{C}$ \cite{Momot2018, Momot:2019lnx, Panasenko2022}. All sensors underwent an automatised optical inspection \cite{Lavrik2022}. There are $1024$ strips  with a pitch of $58\,\mathrm{\mu m}$ on each side; the p-side strips are inclined with respect to the side edge at an angle of $7.5^\circ$. The AC readout pads are located in two rows on the top and bottom of the sensor. Each sensor is connected to two front-end boards (FEBs), each containing eight SMX ASICs, via 32 custom ultra-lightweight aluminium-polyimide microcables up to $500\,\mathrm{mm}$ long with the contribution to the material budget $\lesssim0.3\%X_0$ per module \cite{Protsenko2017}. A picture of the module during its assembling at GSI Detector Laboratory is shown in figure \ref{fig:photo_module}. More details concerning the module structure, construction, calibration and operation can be found in ref. \cite{RodriguezRodriguez2024}.

Several versions of the detector Front-End Boards (FEB) were used during the setup developments: from the single-ASIC FEB-C (visible as a part of the prototype module in figure \ref{fig:msts+e16_sts}, left) to the present date FEB8 (part of the E16 STS module in figure \ref{fig:photo_module}). An alternative implementation of the SMX-based silicon micro-strip detector with few alternative FEB form-factors was done for the \textsc{Strasse} tracker \cite{tuprints19806}.

To achieve the expected noise level during operation, the STS module implements a sophisticated ground and powering scheme. This is an important feature addressed in ref. \cite{Koczon:2020sox}. The HV and ground stability issues are also studied in ref. \cite{Kshyvanskyi2022}.

The Data Acquisition (DAQ) system of the STS detector together with other CBM subsystems feature radiation hard GBTx and Versatile Link based readout: details on the system architecture, protocol and particular implementations are provided in refs. \cite{Lehnert2017} and \cite{onlinecbm2023}.
An alternative DAQ chain with commercial off-the-shelf components based on the GBTx-EMU is described in ref. \cite{Zabolotny2022}

\section{Present and future detector applications}
A few dozens of prototype and pre-series modules has been assembled since 2018 \cite{RodriguezRodriguez2024}. They were involved in multiple laboratory studies and beam tests, primarily with protons at COSY synchrotron (see dedicated chapters of refs. \cite{Rodriguez2020, Pfistner2021}) and heavy-ion collisions in mCBM at SIS18 \cite{MaragotoRodriguez2023}.

\begin{figure}[htbp]
\centering
\includegraphics[height=.351\textwidth]{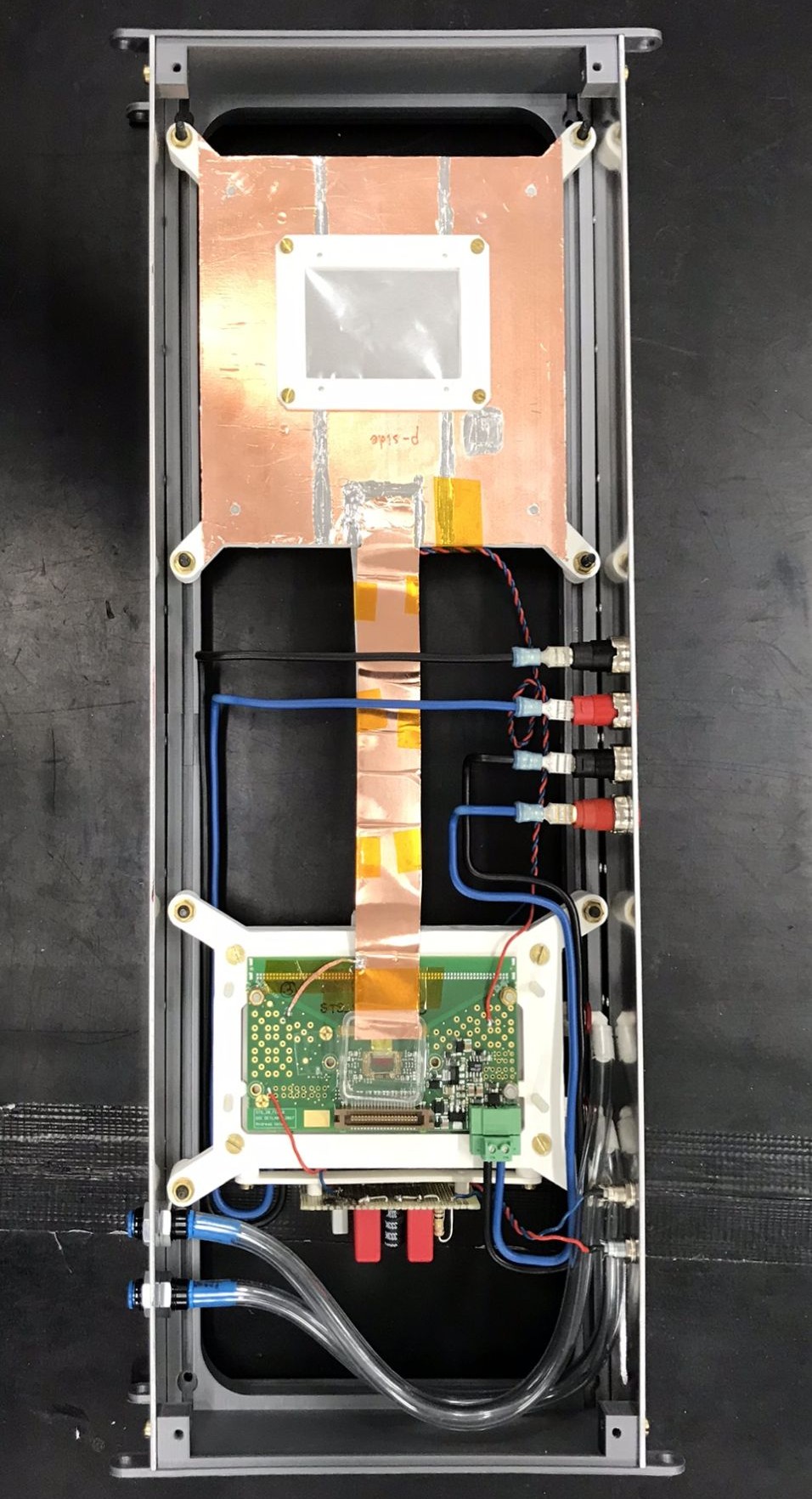}
\qquad
%  [trim={left bottom right top}, clip]
\includegraphics[trim={8cm 0 12cm 0}, clip, height=.35\textwidth]{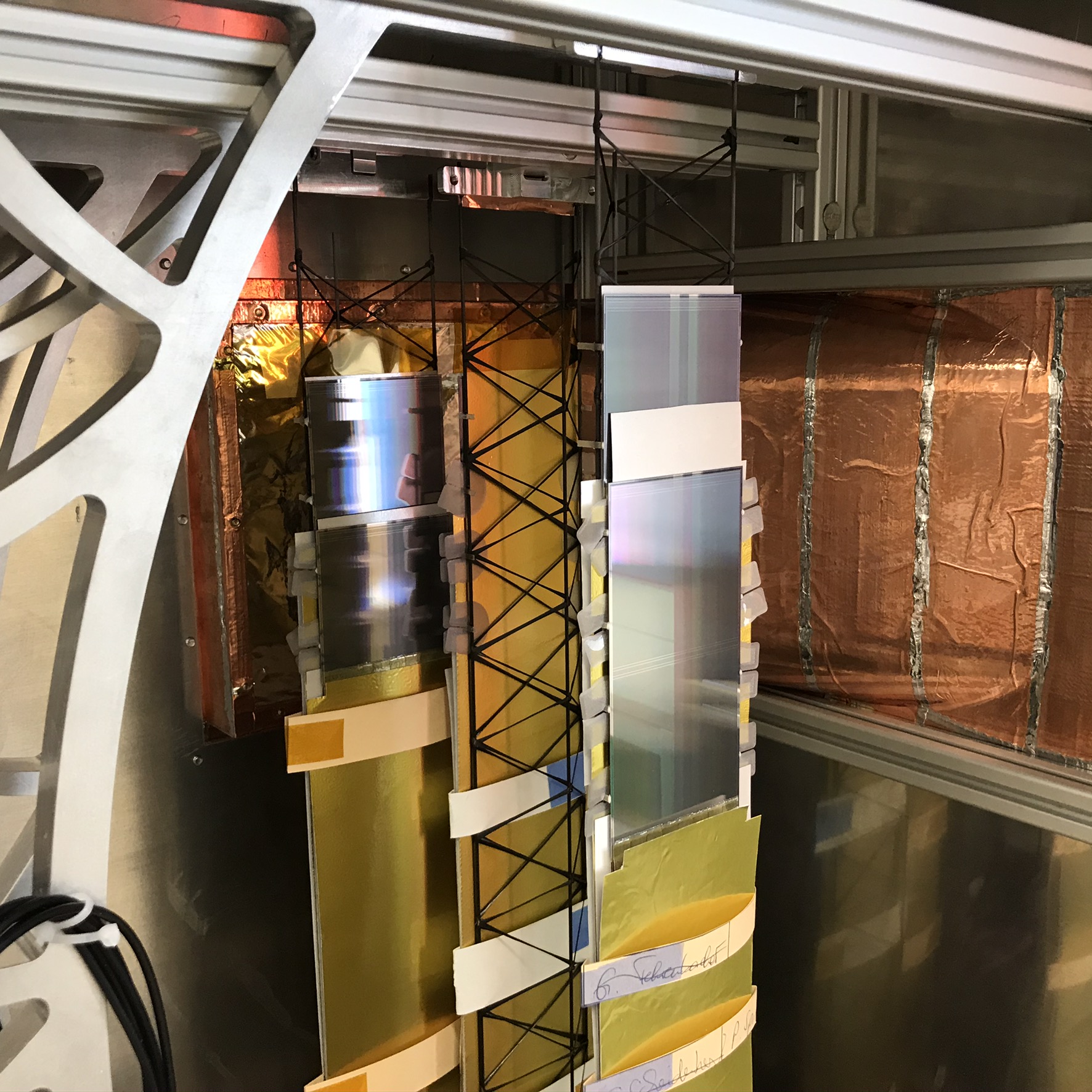}
\qquad
\includegraphics[height=.35\textwidth]{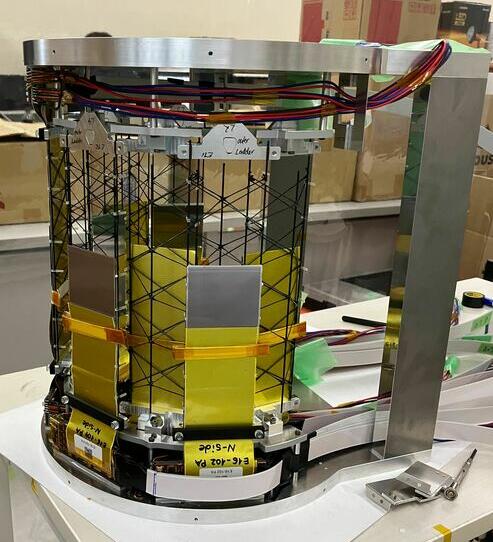}
\caption{\label{fig:msts+e16_sts} Left to right: an early prototype module with two FEB-C prepared for the beam tests at COSY in 2018; mSTS detector during the upgrade in 2021; E16 STS at J-PARC in 2022 (picture from ref. \cite{e16}).}
\end{figure}

The mSTS detector is a fully-integrated functional STS prototype featuring eleven DSDM silicon strip modules of different sensor sizes (see figure \ref{fig:msts+e16_sts}, center); it is a part of the mCBM test setup \cite{CBMCollaboration2017}. The mSTS detector was first proposed in 2017 as a part of the installation of the FAIR Phase\,0 mCBM experiment which aim to prove the free-streaming DAQ and online reconstruction concept of CBM using high rate nucleus-nucleus collisions at SIS18 synchrotron. Since that time the mSTS setup went through two  successful beam campaigns in 2018--2019 and 2020--2022  \cite{GSIHelmholtzzentrumfuerSchwerionenforschung2023}.

The E16 experiment at J-PARC aims to study in-medium modification of the vector mesons $\rho$, $\omega$, and $\phi$, decaying via $e^+e^-$ channel, with a high-intensity $30\,\mathrm{GeV}$ proton beam interacting with a C target and Cu targets at rates up to $40\,\mathrm{MHz}$. It went through three commissioning runs in 2020--2021.
Recently, the silicon strip detector was upgraded by replacing older silicon sensors with new ten STS modules built with  $62\times62\,\mathrm{mm^2}$ sensors (see figure \ref{fig:msts+e16_sts}, right) \cite{Aoki_2023}; it features the GBTx-EMU based DAQ chain \cite{Yamaguchi2022}.

\section{Spatial resolution}

High spacial resolution and detection efficiency are prime objectives of the tracking detectors. According to simulations performed in ref. \cite{Malygina2017}, $\Delta x \lesssim 15\,\mathrm{\mu m}$ and single-hit efficiency $\epsilon_\text{hit}\approx 97\%$ were obtained. Preliminary analysis of the mSTS data supports these estimates.

\section{Timing and rate capabilities }

\begin{figure}[htbp]\centering
 \includegraphics[height=.3\textwidth]{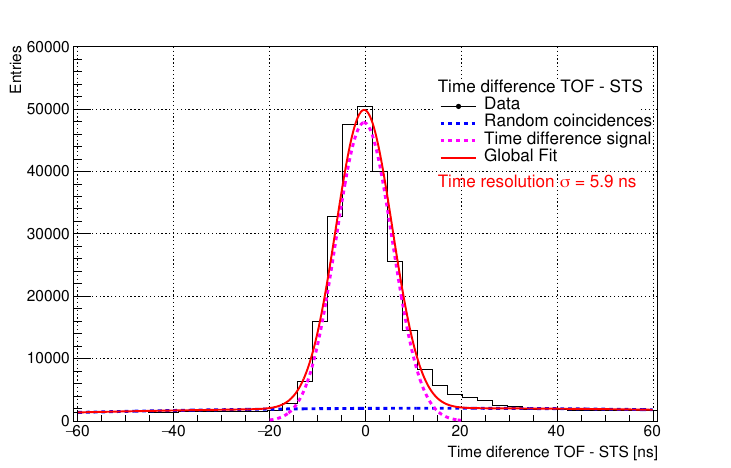}
 \caption{\label{fig:time_res}
 Measured detector time resolution in the mSTS setup  with the time-walk correction applied \cite{MaragotoRodriguez2023}.}
\end{figure}

Timing performance of the STS detector modules was measured in several beam tests with relativistic particles. Time resolution of $\Delta t = 6.9\,\mathrm{ns}$ was obtained with a proton beam of $1.7\,\mathrm{GeV}$ at COSY synchrotron  (see figure \ref{fig:time_res}, left) \cite{Pfistner2021}. Comparable results were obtained with the mCBM data using the prototype Time-of-Flight detector (TOF) as a reference (see figure \ref{fig:time_res}, right) \cite{MaragotoRodriguez2023}.

\begin{figure}[htbp]\centering
 \includegraphics[width=.40\textwidth]{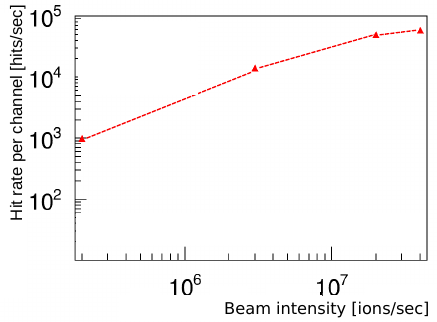}
 \qquad \quad
 \includegraphics[width=.42\textwidth]{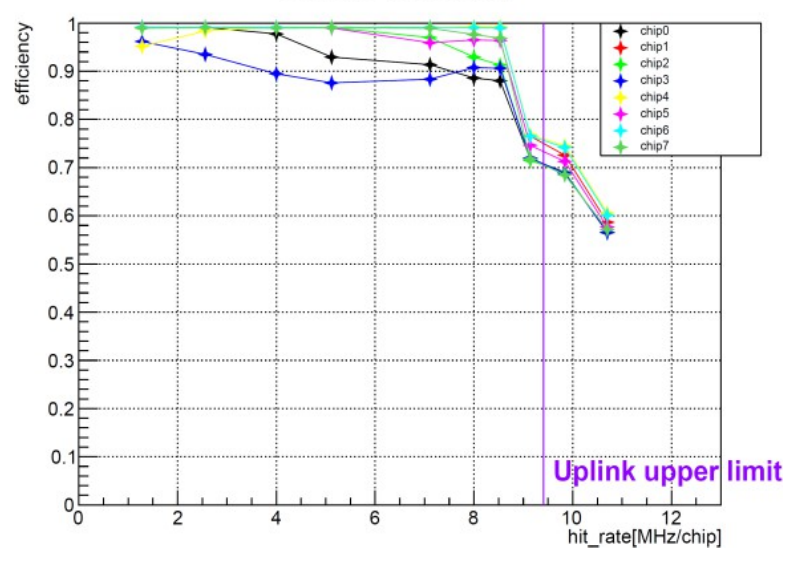}
 \caption{\label{fig:high_rate} Typical hit rate per SMX channel during the mCBM beam test from  ref. \cite{Rodriguez2019}, corrected (left). Detection efficiency per chip measured with ASIC hit generator with E16 STS module; the upper data transfer rate limit of $9.41\,\mathrm{MHz}$ for one uplink (corresponding to one chip) is indicated by the vertical line (right) \cite{Yamaguchi2022}.}
\end{figure}

Several independent studies were conducted to establish the signal rate capabilities of the STS modules: beam tests with mSTS so far showed rate capabilities up to $(128\times40)\,\mathrm{kHz/uplink} = 5.1\,\mathrm{MHz/uplink}$ (figure \ref{fig:high_rate}, left); meantime, E16 team showed stable hit transmission up to $9\,\mathrm{MHz/uplink}$ with the digitally periodically generated   SMX hits (figure \ref{fig:high_rate}, right).

\section{Energy measurements}

Energy loss by a relativistic charge particle penetrating a thin detector follows Landau-Vavilov distribution \cite{Landau:216256, Vavilov:1957zz}.
The width factors $\xi$ and $\sqrt{\delta_2}$, and most probable value $\Delta_p$ are usually used to describe the width and the position of Landau-Vavilov distribution; they both depend on the thickness and material of the detector \cite{Workman2022}. With the mean energy loss of a minimum ionizing particle (MIP) being $\langle\Delta\rangle = 388 \,\mathrm{eV/\mu m}$  we estimate $\Delta_p\approx84\,\mathrm{keV}$ \cite{Workman2022}, $\xi=5.7\,\mathrm{keV}$ \cite{Meroli2011}, and $\sqrt{\delta_2}=7.6\,\mathrm{keV}$\,\cite{Blunck1950ZumES} for MIP in $320\,\mathrm{\mu m}$ silicon.
% https://meroli.web.cern.ch/lecture_StragglingFunction.html
With the mean energy for electron-hole pair production in silicon $\epsilon_\text{Si}=3.64\,\mathrm{eV} $ \cite{Meroli2011}, the most-probable charge deposition is
 $Q_p = \frac{\Delta_p  e}{\epsilon_\text{Si}} = 23.1\times 10^3 \,\mathrm{e} = 3.7\,\mathrm{fC}$.

Taking into account the Poisson distribution for the electron-hole creation and a Fano factor for silicon $F=0.11$ \cite{Rodrigues_2021}, we can finally estimate a typical variation of the charge deposit in the detector:

\begin{equation}\label{eq:eps_int}
 \varepsilon_\text{int} \simeq \frac{e}{\epsilon_\text{Si}} \cdot \frac{  \sqrt{\xi^2 + \delta_2 + F^2 \epsilon_\text{Si}^2  Q_p/e}}{Q_p} =
 \frac{0.25\,\mathrm{fC} \oplus 0.33\,\mathrm{fC} \oplus 0.003\,\mathrm{fC} }{3.7\,\mathrm{fC}} = 11.1\%.
\end{equation}

The SMX chip allows us to measure the charge deposited in the micro-strip silicon sensor with its slow channel and 5-bit flash ADC: thus, assuming only perpendicular tracks resulting in uniquely one-strip clusters and dynamic range of $[0.8,14.0]\,\mathrm{fC}$ the charge uncertainty resulting from the amplitude discretization is about:

\begin{equation}\label{eq:eps_adc}
 \varepsilon_\text{ADC}\simeq \frac{\sigma_\text{ADC}}{Q_p}= \frac{1}{3.7\,\mathrm{fC}}\times \frac{14.0\,\mathrm{fC}-0.8\,\mathrm{fC}}{(2^5-1)\sqrt{12}}=\frac{0.12\,\mathrm{fC}}{3.7\,\mathrm{fC}}=3.3\%,
\end{equation}
which is smaller than uncertainty resulting from the conservative estimate of the effective noise charge (ENC) of $\sigma_\text{ENC}= 1200\,\mathrm{e} \approx 0.2\,\mathrm{fC}$ from ref. \cite{RodriguezRodriguez2024}:
\begin{equation}\label{eq:eps_enc}
 \varepsilon_\text{ENC} \simeq \frac{\sigma_\text{ENC}}{Q_p} = \frac{0.2\,\mathrm{fC}}{3.7\,\mathrm{fC}} = 5.4\%.
\end{equation}
Comparing equations \ref{eq:eps_adc} and \ref{eq:eps_enc} to equation \ref{eq:eps_int} one can see that the measurement precision is dominated by the nature of the interaction rather than by the read-out electronics.

\begin{figure}[htbp]
\centering
\includegraphics[height=.29\textwidth]{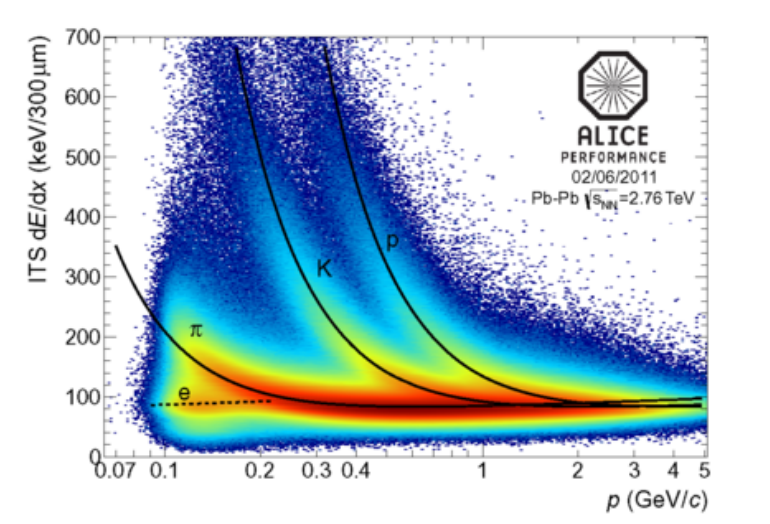}
\qquad
%  [trim={left bottom right top}, clip]
\includegraphics[trim={0 0 0 20pt}, clip, height=.28\textwidth, page=3]{10GeV}
\caption{\label{fig:itc+sts}
Performance studies of the particle identification in the ALICE ITS with Pb+Pb at $\sqrt{s_\mathrm{NN}}=2.76\,\mathrm{TeV}$ (left) \cite{Ivanov2013}. PID in CBM STS with separately simulated thermalized reaction products from Au+Au at $\sqrt{s_\mathrm{NN}}=4.3\,\mathrm{GeV}$ (right) \cite{GSIHelmholtzzentrumFuerSchwerionenforschungGmbH2018}.}
\end{figure}

Particle Identification (PID) using the charged particle energy deposition per unit length of detector material (specific energy loss) is a widely used approach in particle and nuclear physics.  Heavy-ion experiments typically use information about the continuous energy loss in their gaseous trackers \cite{Shao2006, Ivanov2013}. To some extent, this approach is also applicable to silicon tracking detectors, as has been shown with data from the CMS Tracker \cite{Quertenmont2011} and ALICE ITS (see figure \ref{fig:itc+sts}, left) \cite{Ivanov2013, Milano2011}.

Independent measurements of the energy deposit in multiple tracking stations require averaging. As the arithmetic mean does not reduce the uncertainty, CMS Tracker team uses $\langle E\rangle=\left(\Sigma_i 1/E_i^2\right)^{-1/2}$, while ALICE ITC team uses a truncated mean. We obtained the best results using the median energy value; the bands of the median energy deposits for various particles are shown in figure \ref{fig:itc+sts}, right.

\begin{figure}[htbp]
\centering
\includegraphics[width=.35\textwidth]{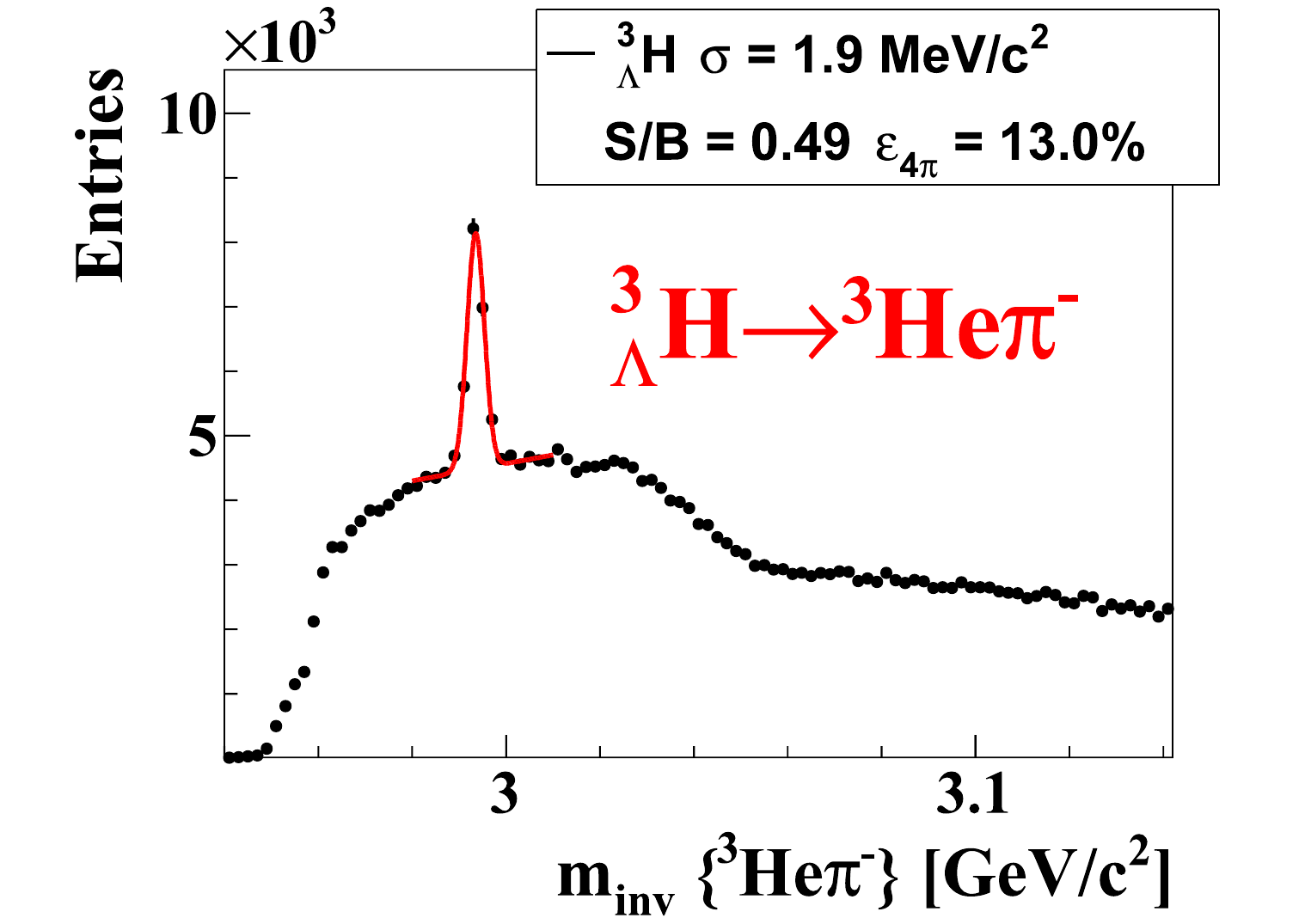}
\qquad \qquad
\includegraphics[width=.35\textwidth]{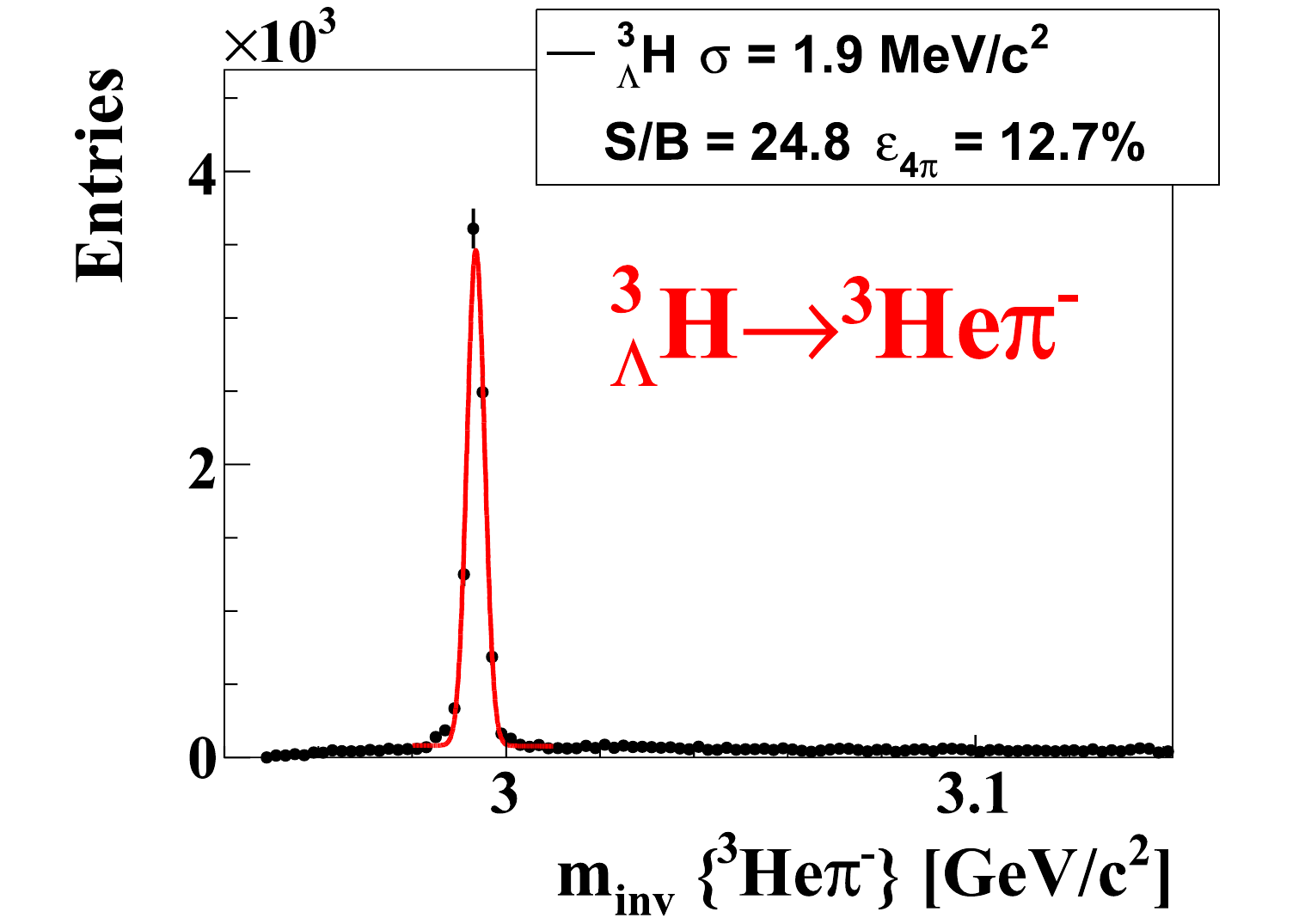}
\caption{\label{fig:hyper} Reconstruction of the $\mathrm{^3_{\Lambda}H}\to \mathrm{^{3}He}\,\pi^-$ decay with TOF only (left) and
with an additional PID from STS (right). Simulation with $5\times10^6$ central
Au+Au collisions at CBM beam energy $10\,A\mathrm{GeV}$ \cite{GSIHelmholtzzentrumFuerSchwerionenforschungGmbH2018}.}
\end{figure}

Applying a simple threshold of $\Delta E/\Delta x > 80\times10^3 \mathrm{e/300\mu m}$ on the median charge deposited in STS detectors results in almost complete separation of double-charged particles from single-charged particles in the STS \cite{Derenovskaya2020}. According to CBM simulations, in the case of hyper tritium decay $\mathrm{^3_{\Lambda}H}\to \mathrm{^{3}He}\,\pi^-$ it helped to increase the signal purity by a factor of 50, as shown in Fig.\ref{fig:hyper}: the main contamination in this channel are misidentified protons and deuterons, which are overlapped with  helium isotopes in TOF detector data \cite{GSIHelmholtzzentrumFuerSchwerionenforschungGmbH2018}.

\acknowledgments

This research would not be possible without the contribution made by the engineers and technical staff of the GSI Detector Laboratory, Module Assembling Team: C. Simons, R. Visinka, O. Bertini, K. Schuenemann and O. Suddia. We are grateful to our colleagues at the Karlsruhe Institute of Technology, who have contributed to the production of the modules: T. Blank, H. Herzenstiel, B. Leyrer. Finally, we thank R. Kappel, I. Keshelashvili, and F. Nickels for their valuable inputs.

% We suggest to always provide author, title and journal data:
% in short all the informations that clearly identify a document.

% \printbibliography
\bibliographystyle{JHEP.bst}
\bibliography{references.bib}

\end{document}